\documentclass[letterpaper,11pt]{article}
\pdfoutput=1
\usepackage{jheppub}
\usepackage{slashed}
\usepackage{graphicx}
\usepackage{subfigure}
\usepackage{soul}
\usepackage{amsmath}
\usepackage{multirow}
\usepackage{tablefootnote}
\usepackage{hyperref}
\usepackage{epstopdf}
\usepackage{lscape}
\usepackage{url}
\usepackage[dvipsnames]{xcolor}
\usepackage[utf8]{inputenc}

\newcommand{\beq}{\begin{equation}}
\newcommand{\eeq}{\end{equation}}
\newcommand{\bea}{\begin{eqnarray}}
\newcommand{\eea}{\end{eqnarray}}
\newcommand{\beqs}{\begin{subequations}}
\newcommand{\eeqs}{\end{subequations}}

\newcommand{\ba}{\begin{array}}
\newcommand{\ea}{\end{array}}

\def\figureautorefname~#1\null{Fig.\,#1\null}
\def\tableautorefname~#1\null{Tab.\,#1\null}

\def\equationautorefname~#1\null{Eq.\,(#1)\null}

\def\m1{M_1}
\def\m2{M_2}
\def\m3{M_3}

\def\ch10{\tilde \chi^0_1}

\def\gev{\,{\rm GeV}}

\def\to{\rightarrow}

\newcommand{\lsim}{\mathrel{\mathop{\kern 0pt \rlap
  {\raise.2ex\hbox{$<$}}}
  \lower.9ex\hbox{\kern-.190em $\sim$}}}
\newcommand{\gsim}{\mathrel{\mathop{\kern 0pt \rlap
  {\raise.2ex\hbox{$>$}}}
  \lower.9ex\hbox{\kern-.190em $\sim$}}}

\definecolor{pink}{RGB}{255,105,180}

\def\cosba{\cos(\beta-\alpha)}

\newcommand{\bpm}{\begin{pmatrix}}
\newcommand{\epm}{\end{pmatrix}}

\newcommand{\tanb}{\tan \beta}

\allowdisplaybreaks[4]

\title{Exploring the low $\tanb$ region of two Higgs doublet models at the LHC}

\author[a]{Wei Su}
\author[a]{, Martin White}
\author[a]{, Anthony G. Williams}
\author[b]{ and Yongcheng Wu}
\preprint{ADP-19-20/T1100}

\affiliation[a]{ARC Centre of Excellence for Particle Physics at the Terascale, Department of Physics,University of Adelaide, South Australia 5005, Australia}

\affiliation[b]{Ottawa-Carleton Institute for Physics, Carleton University, 1125 Colonel By Drive, Ottawa, Ontario K1S 5B6, Canada}

\emailAdd{wei.su@adelaide.edu.au, anthony.williams@adelaide.edu.au, martin.white@adelaide.edu.au, ycwu@physics.carleton.ca}

\abstract{ 
Current interpretations of the LHC results on two Higgs doublet models (2HDM) underestimate the sensitivity due to neglecting higher order effects. In this work, we revisit the impact of these effects using the current cross-section times branching ratio limits of the $A\to hZ, H \to VV$ and $H\to hh$ channels. With a degenerate heavy Higgs mass $m_\Phi$, we find that the LHC searches gain sensitivity to the small $\tanb$ region after including loop corrections, even close to $\cosba=0$ which is not reachable at tree level for all types of 2HDM. For a benchmark point with $m_\Phi=300$ GeV, $\tanb<1.8(1.2)$ can be probed for the Type-I(II) 2HDM model for $\cosba=0$. When the deviation from $\cosba=0$ is larger, the region for which current searches have exclusion potential becomes larger.}


\begin{document}
\maketitle
\flushbottom

\section{Introduction and motivation}
\label{sec:intro}

The discovery of a Standard Model (SM)-like Higgs boson at the Large Hadron Collider (LHC)~\cite{Aad:2012tfa,Chatrchyan:2012xdj} strongly informs LHC searches for physics beyond-the-SM (BSM), especially for an expanded Higgs sector. Motivations for an extension of the SM arise from both theoretical and observational considerations~\cite{Giudice:2008bi}.

Two Higgs doublet models (2HDMs) are well-motivated scenarios that provide the simplest generalization of the SM Higgs sector~\cite{Branco:2011iw}. After electroweak symmetry breaking (EWSB), the general 2HDM will generate 5 mass eigenstates, a pair of charged Higgs boson $H^\pm$, one CP-odd Higgs boson $A$ and two CP-even Higgs bosons, $h, H$. Here we take the lighter $h$ as the observed SM-like Higgs. The spectrum can be studied through direct heavy Higgs searches at hadron colliders, or indirect, precision measurements of the couplings of the SM-like Higgs bosons at the LHC or future lepton colliders~\cite{Aaboud:2017sjh,CMS-PAS-HIG-17-020,Aaboud:2017gsl,Aaboud:2017rel,Sirunyan:2018qlb,Aaboud:2017yyg,Aaboud:2017cxo,Aaboud:2018eoy,Khachatryan:2016are,Aaboud:2018ftw,Sirunyan:2018iwt,ATLAS-CONF-2016-089,Aaboud:2018gjj,CMS-PAS-HIG-16-031,Chen:2018shg,Bian:2017gxg}. Direct search signals are a fruitful area of study because of the interesting variety of heavy Higgs decay channels $H/H^\pm/A \to f\bar f^{'}, VV^{'}, Vh, hh$. There are also some exotic decay channels such as $A/H \to H/A Z$~\cite{Kling:2018xud,Kling:2016opi,Li:2015lra}. So far, the assessments of the null results at the LHC include only the tree-level Higgs couplings, under the assumption that higher-order corrections are not important for the interpretation of the current LHC datasets. However, this neglects the fact that, in some special regions, loop corrections can play a key role. For example, in the $A\to Zh,\ H\to VV$, and $ H\to hh$ channels, the tree-level couplings are proportional to $\cosba$. They therefore vanish in the ``alignment limit'' of $\cosba=0$ and, as a result, give no constraint around the $\cosba=0$ region~\cite{Aaboud:2017cxo, Sirunyan:2019xls, Sirunyan:2018qlb,Aaboud:2017rel,Aaboud:2018bun,Aaboud:2017gsl,CMS:2019kjn,Aad:2019uzh,Sirunyan:2018two}. Loop corrections change this picture substantially, however, and we will find below that with a combination of these channels the region around $\cosba=0$ is no longer unreachable. Our study shows that the searches in these channels are sensitive to small $\tanb$ values even for $ \cosba=0$ which is unconstrained at tree level, and we present the updated limits on the parameters of two Higgs doublet models for degenerate heavy Higgs masses. The study shows that the small $\tanb$ region can be strongly constrained for all four types of 2HDM, where the experimental limits are applicable.

The paper is structured as follows. In \autoref{sec:2hdm}, we give a brief introduction to 2HDMs, summarizing the experimental and theoretical results for the four decay channels described above, with the detailed formulae at one-loop level given in \autoref{app:Couplings}. We present our individual channel analyses as well as the combined results in \autoref{sec:result}. Finally we give our main conclusions in \autoref{sec:con}.

\section{Two Higgs doublet models}
\label{sec:2hdm}
\subsection{2HDM Higgs sector}

For the general CP-conserving 2HDM, there are two ${\rm SU}(2)_L$ scalar doublets $\Phi_i\ (i=1,2)$ with hyper-charge $Y=+1/2$,
\begin{equation}
\Phi_{i}=\begin{pmatrix}
  \phi_i^{+}    \\
  (v_i+\phi^{0}_i+iG_i)/\sqrt{2}
\end{pmatrix}\,.
\end{equation}
where $v_i$ are the vacuum expectation values (vev) of the two doublets after EWSB with $v_1^2+v_2^2 = v^2 = (246\ {\rm GeV})^2$ and $\tanb\equiv v_2/v_1$.

The Higgs sector Lagrangian of the 2HDM can be written as
\begin{equation}\label{equ:Lall}
\mathcal{L}=\sum_i |D_{\mu} \Phi_i|^2 - V(\Phi_1, \Phi_2) + \mathcal{L}_{\rm Yuk}\,,
\end{equation}
with a Higgs potential of
\begin{eqnarray}
\label{eq:L_2HDM}
 V(\Phi_1, \Phi_2) &=& m_{11}^2\Phi_1^\dag \Phi_1 + m_{22}^2\Phi_2^\dag \Phi_2 -m_{12}^2(\Phi_1^\dag \Phi_2+ h.c.) + \frac{\lambda_1}{2}(\Phi_1^\dag \Phi_1)^2 + \frac{\lambda_2}{2}(\Phi_2^\dag \Phi_2)^2  \notag \\
 & &+ \lambda_3(\Phi_1^\dag \Phi_1)(\Phi_2^\dag \Phi_2)+\lambda_4(\Phi_1^\dag \Phi_2)(\Phi_2^\dag \Phi_1)+\frac{\lambda_5}{2}   \Big[ (\Phi_1^\dag \Phi_2)^2 + h.c.\Big]\,,
\end{eqnarray}
where we have assumed $CP$ conservation, and a soft $\mathbb{Z}_2$ symmetry breaking term $m_{12}^2$.

After EWSB, three Goldstone bosons are eaten by the SM gauge bosons $Z$, $W^\pm$, providing their masses. The remaining physical mass eigenstates are $h, H, A$ and $H^\pm$.  The usual eight parameters appearing in the Higgs potential are $m_{11}^2, m_{22}^2, m_{12}^2, \lambda_{1,2,3,4,5}$, and can be transformed to a more convenient choice of the mass parameters: $v, \tan\beta, \alpha, m_h, m_H, m_A, m_{H^\pm}, m_{12}^2$, where $\alpha$ is the rotation angle diagonalizing the CP-even Higgs mass matrix.

$\mathcal{L}_{\rm Yuk}$ in the Lagrangian represents the Yukawa interactions of the two doublets. To avoid tree-level flavor-changing neutral currents (FCNC), all fermions with the same quantum numbers are made to couple to the same doublet~\cite{Glashow:1976nt,Paschos:1976ay}. By assigning different doublets to different fermions, in general, there are four possible types of Yukawa coupling: Type-I, 
Type-II,
Type-LS (Lepton specific),
and Type-FL (Flipped).
In this work, the most relevant Yukawa couplings are $y_{b}$ and $y_t$, which matter for the $A\to Zh$ with $h\to b\bar b$ process and constitute the dominant loop corrections. Thus we will focus on the Type-I and Type-II 2HDMs. The situation in the Type-LS and Type-FL 2HDMs would be similar:
\begin{align}
\kappa_b\equiv\frac{y_{b}^{\rm 2HDM}}{y_{b}^{\rm SM}}=\begin{cases}
\frac{c_\alpha}{s_\beta} & \text{Type-I, LS}\\
-\frac{s_\alpha}{c_\beta} & \text{Type-II, FL}
\end{cases}, \quad \kappa_t \equiv \frac{y_t^{\rm 2HDM}}{y_t^{\rm SM}} = \begin{cases}
\frac{c_\alpha}{s_\beta} & \text{Type-I, LS}\\
\frac{c_\alpha}{s_\beta} & \text{Type-II, FL}.
\end{cases}
\end{align}
Here we take $c_x = \cos x$ and  $s_x= \sin x$.

%

\subsection{Heavy Higgs Decay Channels}
\label{sec:decay}
The heavy Higgs decay channels we are interested in and the corresponding tree-level couplings are
\begin{align}
&{\bf A\to Zh} & &g_{AhZ} = \frac{m_Z}{v}c_{\beta-\alpha}(p_A^\mu - p_h^\mu) \label{eq:c_AhZ},\\
&{\bf H\to ZZ} & &g_{HZZ} = \frac{2m_Z^2}{v}c_{\beta-\alpha} \label{eq:c_HZZ}, \\
&{\bf H\to WW} & &g_{HWW} = \frac{2m_W^2}{v}c_{\beta-\alpha}, 
\label{eq:c_HWW}\\
&{\bf H\to hh} & &g_{Hhh} = -\frac{c_{\beta-\alpha}}{4s_{2\beta}v}\left(\frac{4m_{12}^2}{s_\beta c_\beta}(3s_\alpha c_\alpha-s_\beta c_\beta)-2(2m_h^2+m_H^2)s_{2\alpha}\right).
\label{eq:c_Hhh}
\end{align}
At the tree-level, all of these couplings vanish in the alignment limit where $c_{\beta-\alpha} = 0$. As a consequence, all are currently thought to be irrelevant for constraining 2HDMs in the tree-level alignment limit. 

However, at one-loop level, the definition of the alignment limit will shift channel-by-channel from the previous definition $c_{\beta-\alpha} = 0$.
For $c_{\beta-\alpha} = 0$ and $m_{H} = m_{A} = m_{H^\pm} = \sqrt{(m_{12}^2/s_\beta c_\beta)}\equiv m_{\phi}$, the one-loop coupling expressions for the relevant vertices can be  simplified and given by

\begin{align}
C_{hAZ} &\approx \frac{3e^3(p_h^\mu-p_A^\mu)}{64\pi^2s_W^3c_Wm_W^2}\sum_{f=t,b}\xi_fm_f^2\times\Big(\mathcal{PV}_1\Big)\label{eq:loop-haz},\\
C_{HWW} &\approx \frac{3e^3}{32\pi^2s_W^3m_W}\sum_{f=t,b}\xi_fm_f^2\times\left({g^{\mu\nu}\Big(\mathcal{PV}_2\Big)} + {\frac{p_1^\mu p_2^\nu}{m_W^2}\Big(\mathcal{PV}_3\Big)}\right), \\
C_{HZZ} &\approx \frac{3e^3g^{\mu\nu}}{8\pi^2s_W^3c_W^2m_W}\sum_{f=t,b}\xi_fm_f^2\times\left(\left(\left(c_L^f\right)^2+\left(c_R^f\right)^2\right)\times\left({g^{\mu\nu}\Big(\mathcal{PV}_4\Big)} + {\frac{p_1^\mu p_2^\nu}{m_Z^2}\Big(\mathcal{PV}_5\Big)}\right)\right.\nonumber \\
&\quad \left. +c_L^fc_R^f\times\left({g^{\mu\nu}\Big(\mathcal{PV}_6\Big)} + {\frac{p_1^\mu p_2^\nu}{m_Z^2}\Big(\mathcal{PV}_7\Big)}\right)\right),\\
C_{Hhh} &\approx \frac{3e^3}{32\pi^2s_W^3m_W^3}\sum_{f=t,b}\xi_fm_f^2\times\Big(\mathcal{PV}_8\Big),
\label{eq:loop_Hhh}
\end{align}
where $\xi_t = \cot\beta$ for both the Type-I and Type-II models, while $\xi_b = \cot\beta$ for the Type-I model and $\xi_b = -\tan\beta$ for the Type-II model. $c_L^f=I_f - Q_f s_W^2$ and $c_R^f=-Q_f s_W^2$ are the couplings between fermions and the $Z$-boson. $(\mathcal{PV}_i)$ represents some combination of Passarino-Veltman functions~\cite{Passarino:1978jh} which only depends on the masses. The full expressions in this limit can be found in~\autoref{app:Couplings}.

In the tree-level alignment limit, the dominant contributions to the above couplings come from the fermion (top and/or bottom) loop. Thus, all of these couplings are related to the Yukawa coupling modifier $\xi_f$. In both the Type-I and Type-II models, $\xi_t = \cot \beta$ which will be significantly enhanced in the low $\tan\beta$ region. Together with the large value of $m_t$, the unexplored region around $c_{\beta-\alpha}=0$ at low $\tan\beta$ can thus be probed by these channels.

We calculated the complete expressions for the amplitudes for all decay channels ($ff$, $VV$, $SV$ and $SS$) of all scalars ($h, H, A, H^\pm$) using {\tt FeynArts}~\cite{Hahn:2000kx} and {\tt FormCalc}~\cite{Hahn:2016ebn}, using the 2HDM model files including full 1-loop counter terms and renormalization conditions. All renormalization constants are determined using the on-shell renormalization scheme, except $m_{12}^2$ which is renormalized by $\overline{\rm MS}$~\cite{Chen:2018shg}. Note that, for the $SVV$ type couplings, we also include the Lorentz structure $p_1^\mu p_2^\nu/m_V^2$ in the calculation which does not appear in the tree-level calculation. The $\epsilon_{\mu\nu\rho\sigma}$ structure which represents a CP-odd interaction between the scalar and vector bosons can be safely ignored in our calculation, since for the CP-even scalar $H$, the presence of such structure indicates CP violation which can only come from the CP phase in the CKM matrix in the CP conserving 2HDM. All of these NLO amplitudes are then implemented in {\tt 2HDMC}~\cite{Eriksson:2009ws} to fully determine the branching fractions of different channels and the total width of each particle. 

\subsection{Heavy Higgs search results at LHC Run-II}
\label{sec:LHC_reports}
A variety of searches for heavy Higgs bosons have been conducted by the ATLAS and CMS collaboration. Here we use the published cross-section times branching ratio limits to directly constrain the 2HDM parameter space, with the {\tt SusHi} package \cite{Liebler:2016ceh} for the production cross-section at NNLO level, and our own improved {\tt 2HDMC} code, which adds loop-level effects to the public {\tt 2HDMC} code~\cite{Eriksson:2009ws}, for the branching ratios. For each Higgs production and decay process of interest, there exist both ATLAS and CMS public results, and these are non-trivial to apply in practice due to the assumption of a particular width in the presentation of the final results. Since we are not modelling a continuous likelihood, we do not combine these results but, instead, take the most constraining, or the limit with the widest region of applicability from the perspective of the Higgs boson widths. For points where the heavy Higgs decay widths are larger than those assumed for the derivation of the published limits, we use the largest available $\Gamma_H/m_H$ limit, but overlay plots with a region of applicability to indicate the need for caution in our reinterpretations (a device borrowed from experimental reports such as Fig.7 of \cite{Sirunyan:2019xls}). The analyses that we consider include the following:


\begin{enumerate}
  \item \textbf{$A\to Zh$:} Both the ATLAS~\cite{Aaboud:2017cxo} and CMS~\cite{Sirunyan:2019xls} collaborations have presented results with $h\to b\bar b$. Here we choose the ATLAS report for reinterpretation, due to the clear statement of decay width $\Gamma_A/m_A \leq 10\%$, and it includes both $b$-associated and gluon fusion production modes.
  \item \textbf{$H\to ZZ$:} We choose the CMS report~\cite{Sirunyan:2018qlb} for reinterpretation rather than the ATLAS report~\cite{Aaboud:2017rel,Aaboud:2018bun}, since it has a clear $\Gamma_H/m_H$ table with an upper limit of $\Gamma_H/m_H=$30\%. The report uses all the heavy Higgs production modes at CMS.
  \item \textbf{$H\to WW$:} We use the ATLAS results~\cite{Aaboud:2017gsl} rather than the CMS~\cite{CMS:2019kjn} results, since the ATLAS collaboration has published limits on $\sigma\times \rm Br$ in terms of different $\Gamma_H/m_H$ assumptions (the narrow width approximation which assumes $\Gamma_H/m_H=$2\%, 5\%, 10\% and 15\%). This ATLAS report only consider the gluon fusion production mode for heavy Higgs.
  \item \textbf{$H\to hh$:} We use the ATLAS~\cite{Aad:2019uzh} results which  have a detailed description of the $\Gamma_H/m_H$ assumptions. The results involve a combination of the SM-like Higgs decay channel results, combining $b\bar b \gamma\gamma$, $b\bar bb\bar b$, and $b\bar b \tau^+\tau^-$ for $\Gamma_H/m_H \leq 2\%$; $ b\bar bb\bar b$ and $b\bar b \tau^+\tau^-$ for $2\% \geq \Gamma_H/m_H \leq 5\%$; and $  b\bar b$ and $\tau^+\tau^-$ for $5\% \leq \Gamma_H/m_H \leq 10\%$. As in the $H\to ZZ$ case, the heavy Higgs production here is assumed to include all modes. 
  In the report, the cross section times branching ratio limits are estimated based on the assumption that all SM-like Higgs couplings, except for the triple Higgs coupling itself $g_{hhh}$, are the same as those expected in the SM. We use the acceptance and efficiency information given and then rescale to get the cross section times branching ratio limits for the 2HDM where the SM-like Higgs couplings can depart from their SM values. The CMS results can be found in \cite{Sirunyan:2018two}. 
\end{enumerate}

\section{Results}
\label{sec:result}
To build intuition for the full impact of the one-loop corrections to heavy Higgs decays, we first provide a detailed comparison of the tree-level and one-loop-level results for each channel separately in the $\cosba - \tanb$ plane.

\subsection{The $A\to Zh (h \to b\bar b)$ channel}

\begin{figure}[!tb]
\begin{center}
\includegraphics[width=14.5cm]{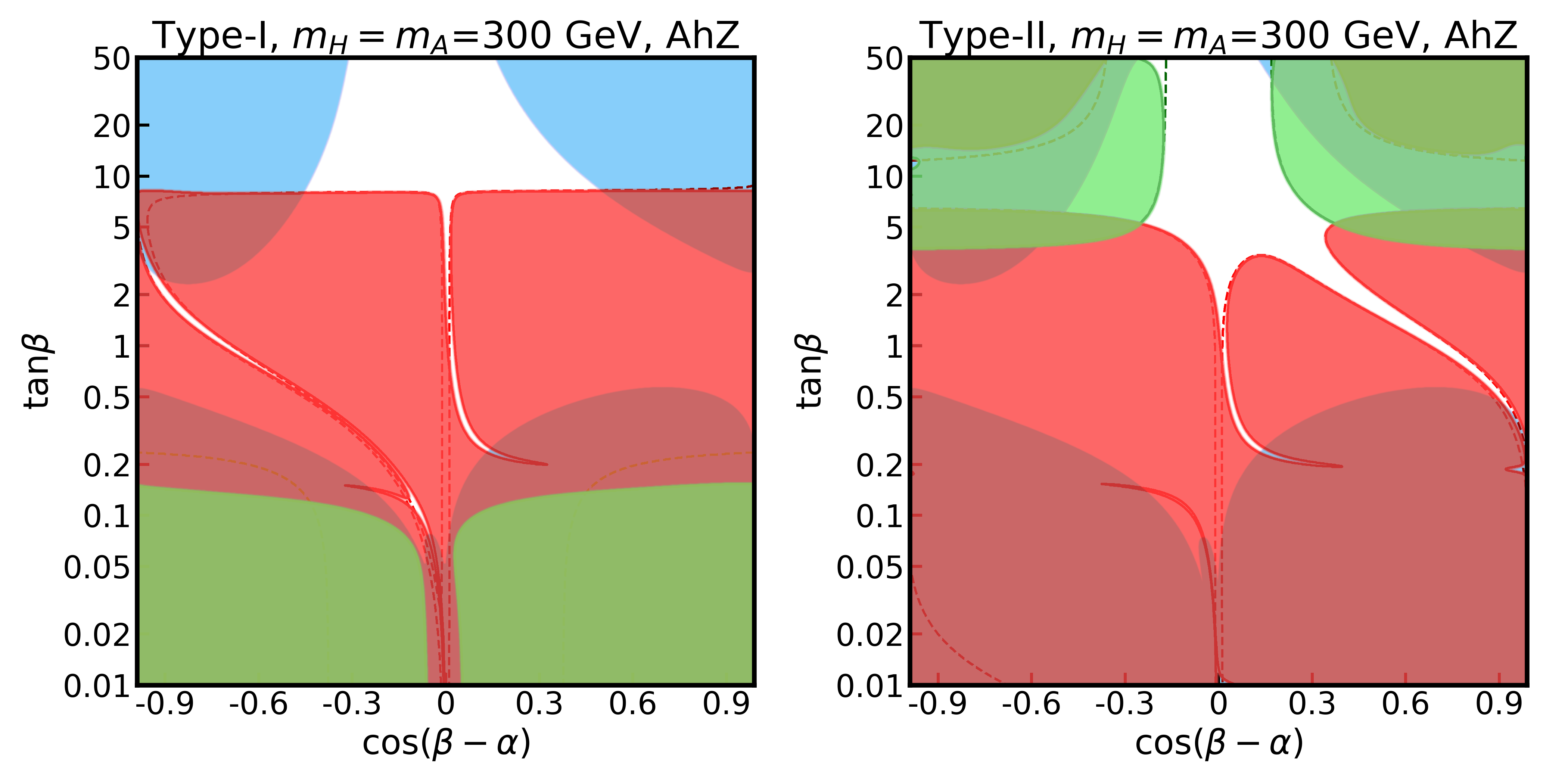}
\caption{Regions in the $\cosba$-$\tan \beta$ plane excluded by experimental results in the $A\to Zh$ channel in the Type-I (left panel) and Type-II (right panel) 2HDM. We have assumed degenerate heavy Higgs masses of 300 GeV. The dashed lines in the figures reproduce the tree-level constraints, while the green and red regions are one-loop level results. The green ones are assumed to be excluded by the $b$-associated production channel, whilst the red regions show the excluded regions by the gluon-fusion production channel. The blue backgrounds represent the points with $\Gamma_A/m_A > 10\%$ or $\Gamma_H/m_H > 10\%$ at one-loop level. }
 \label{fig:AhZ_tanbcba}
\end{center}
\end{figure}

In~\autoref{fig:AhZ_tanbcba}, the constrained parameter space is shown in the $\cosba - \tanb$ plane, with the benchmark point $m_A=m_H=m_{H^\pm} = 300 \gev$, $m_{H}^2=m_ {12}^2/(s_\beta c_\beta)$ . 
The left panel is for the Type-I 2HDM and the right one is for the Type-II. The results are shown separately for the $b$-associated and gluon fusion production modes, as the green and red shadow regions respectively. In the figures the tree-level results are shown with dashed lines. 
For the Type-I 2HDM, the effects of the limits within the region of applicability can exclude the $\tanb < 8$ region, except for two bands. The central band around $\cosba=0$ has a small tree-level $AhZ$ coupling as in \autoref{eq:c_AhZ} and the lower left curve band has a small $hbb$ coupling. The green and red regions represent the one-loop level results excluded by the $b$-associated production and gluon-fusion production channel respectively. We display the regions of $\Gamma_A/m_A >0.1$ or $\Gamma_H/m_H >0.1$ with light blue backgrounds. Generally $\tanb<8$ is also strongly constrained, but the loop-level effects shift the the allowed region around $\cosba=0$ in the small $\tanb$ region. The $0.2< \tanb <2$ allowed region is shifted to the right whilst the $\tanb<0.2$ region is shifted to left. For the Type-II 2HDM, the shifted allowed region at small $\tanb$ is similar to that seen in the Type-I scenario. The allowed band for $\cosba>0.3$ arises because the $hbb$ coupling becomes small in that region.  Meanwhile the constraints in large $\tanb$ region are quite different because of the effect of the $hbb$ Yukawa couplings, which affect the $b$-associated production cross-section. 

\subsection{The $H \to VV$ channel}

As shown in \autoref{eq:c_HZZ}, and \autoref{eq:c_HWW}, the tree-level $HVV$ couplings are only proportional to $\cosba$, and they are therefore independent of the 2HDM model type. 
At one-loop level, the couplings will become type-dependent through fermion corrections. However, the main difference comes from the production which is similar to the $A\to hZ$ case. Hence, we will only show the results for one type and briefly comment on the difference after that. 

\begin{figure}[tb]
\begin{center}
\includegraphics[width=14.5cm]{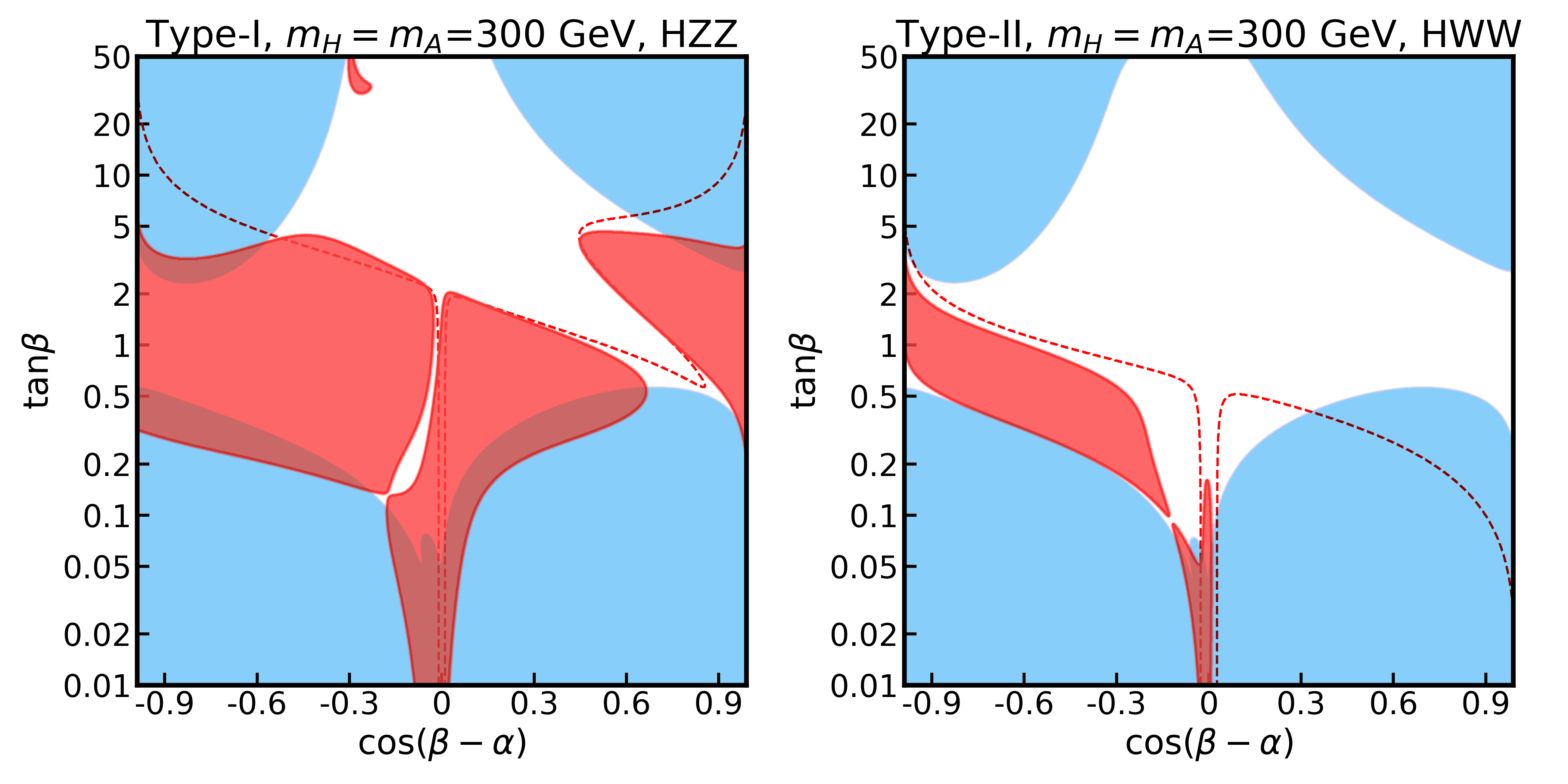}
\caption{Regions in the $\cosba$-$\tan \beta$ plane excluded by experimental results in the $H \to VV$ channel, with assumed degenerate heavy Higgs masses of 300 GeV. The red dashed lines in the figures represent the tree-level constrained regions, while the red shadow regions are excluded by the one-loop-level results. In the case of diboson decay, the four types are exactly the same at tree-level and the loop effects do not differ significantly for different types. Here the left panel is for the Type-I 2HDM $H \to ZZ$ channel and the right is for the Type-II $H\to WW$ channel. The blue backgrounds represent the regions where $\Gamma_A/m_A > 10\%$ or $\Gamma_H/m_H > 10\%$ at one-loop level.}
 \label{fig:Hvv_tanbcba}
\end{center}
\end{figure}


In \autoref{fig:Hvv_tanbcba}, we show the constrained parameter space in the $\cosba - \tanb$ plane for the Type-I 2HDM $H \to ZZ$ channel (left panel), and for the Type-II $H\to WW$ channel (right panel), with the benchmark point $m_A=m_H=m_{H^\pm} = 300 \gev$, $m_{12}^2/(s_\beta c_\beta)=m_H^2$. For the tree-level results shown with dashed red lines, in the region the limits are applicable, $\tanb<2(1)$ is strongly constrained for the Type-I and Type-II 2HDMs except for the central band around the $\cosba=0$ region.
At the one-loop level, the excluded regions represented by red shadow are largely separated into two parts.  For the $H\to ZZ$ channel, one excluded part is at $0.3 < \tanb <5, \cosba<0$ and the other one has $\tanb<2$ around $ \cosba=0$. For the Type-II $H\to WW$ scenario, one is around $\tanb=1$ with $\cosba<0$ and the other one has $\tanb<0.2$ around $ \cosba=0$. 
The large deviations from the tree-level results in the low $\tan\beta$ region are mainly from the influence of the large triple scalar couplings which give rise to large corrections through Higgs field renormalization as well as the large branching ratio of $H\to hh$. 
Similar to \autoref{fig:AhZ_tanbcba}, the regions of $\Gamma_A/m_A >0.1$ or $\Gamma_H/m_H >0.1$ are displayed with light blue backgrounds.

The large $\tanb$ region is less constrained because the production cross-sections of the heavy Higgs bosons are much smaller. For the $H \to ZZ$ channel, the report from CMS~\cite{Sirunyan:2018qlb} includes both gluon fusion and $b$-associated production. As a result, the different types of exclusion limit for the $HZZ$ channel are different at large $\tanb$ to the case of the $A\to Zh$ channel. On the other hand, results are only reported for gluon-fusion production in the case of the $HWW$ channel. Hence the results are quite similar for the different model types. We also note the small red region at large $\tanb$, which comes from the noncontinuous cross section times branching ratio limits in that region.  

\subsection{The $H \to hh$ channel}

As shown in \autoref{eq:c_Hhh}, the $Hhh$ couplings at tree-level are type-independent. 
At one-loop level, though, the correction is type-dependent. The main differences between types come from the production mode as in previous cases. At large $\tan\beta$, $b$-associated production makes a big difference for different types.

\begin{figure}[tb]
\begin{center}
\includegraphics[width=14.5cm]{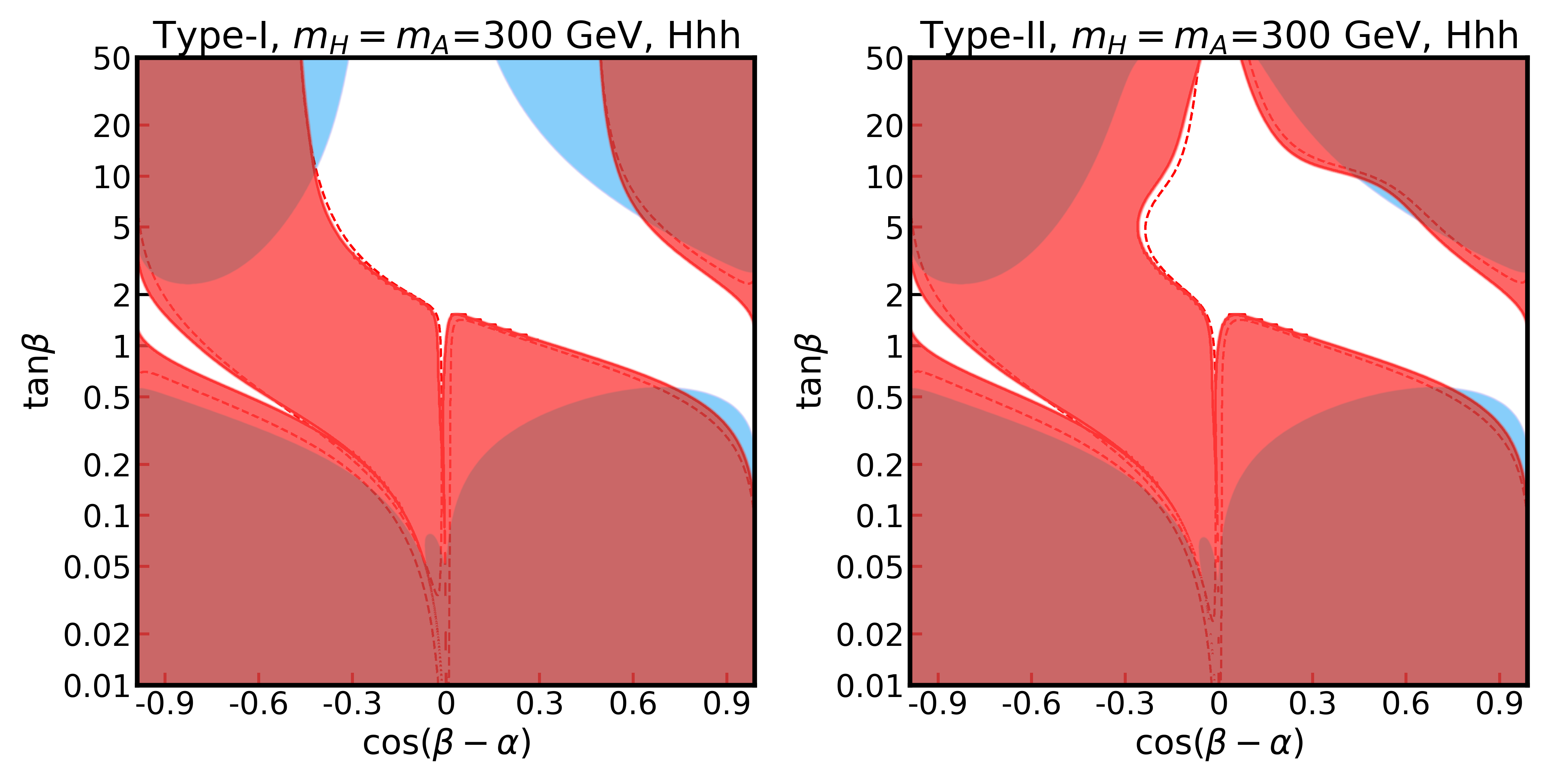}
\caption{Study of the impact of measurements in the $H \to hh$ channel in the $\cosba$-$\tan \beta$, with degenerate heavy Higgs masses of 300 GeV.  As before the dashed red lines in the figure show results based on tree-level calculations, and the excluded region at loop level are red. The left plot is for the Type-I 2HDM and the right for Type-II. Here the Higgs production modes include both gluon fusion and $b$-associated production, and the latter provides the main difference between the two model types. The blue backgrounds represent the points with $\Gamma_A/m_A > 10\%$ or $\Gamma_H/m_H > 10\%$ at one-loop level.}
 \label{fig:Hhh_tanbcba}
\end{center}
\end{figure}
The results for the $H\to hh$ channel are shown in \autoref{fig:Hhh_tanbcba}, where the left panel is for Type-I and the right panel is for Type-II, with the dashed red lines for tree-level results and red regions for loop-level results. Here the benchmark point is still $m_A=m_H=m_{H^\pm} = 300 \gev$, $m_{12}^2/(s_\beta c_\beta)=m_H^2$. For both of the Yukawa types, there are two allowed regions. One is the region around $\cosba=0$, and the other one is the band starting from $\tanb=1.5, \cosba=-1$ to $\tanb=0.01, \cosba=0$. The main feature here is that, in the parameter space where reports limits are applicable (non-blue region), there is nearly no allowed region, especially for $\tan<0.3$, at one-loop level, which is still allowed at tree-level.

%

\subsection{Loop effects summary}

Our individual analyses of the channels $A\to hZ, H\to VV/hh$ have revealed that loop effects can contribute greatly in some regions, especially the small $\tanb$ region. Now we display the combined results with the same benchmark point $m_A=m_H=m_{H^\pm} = 300 \gev$, $m_{12}^2/(s_\beta c_\beta)=m_H^2$.

\begin{figure}[tb]
\begin{center}
\includegraphics[width=14.5cm]{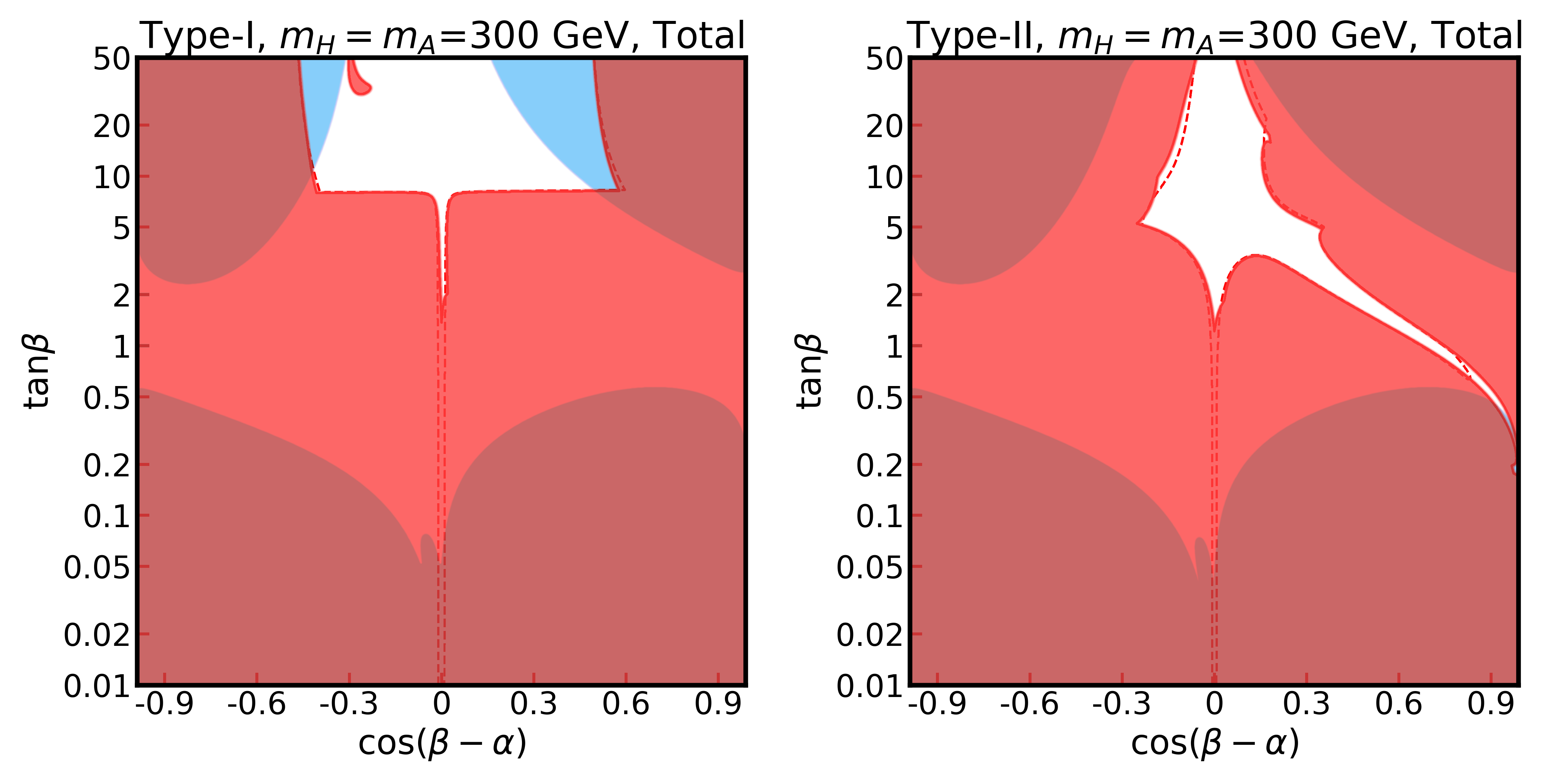}
\caption{Combined study in the $\cosba$-$\tan \beta$ plane, with assumed degenerate heavy Higgs masses of 300 GeV. The dashed lines represent tree-level results and the red represent loop-level constrained regions. The left (right) is for Type-I (Type-II) 2HDM. The blue background represent the points with $\Gamma_A/m_A > 10\%$ or $\Gamma_H/m_H > 10\%$ at one-loop level.}
 \label{fig:total_tanbcba}
\end{center}
\end{figure}

\begin{figure}[tb]
\begin{center}
\includegraphics[width=14.5cm]{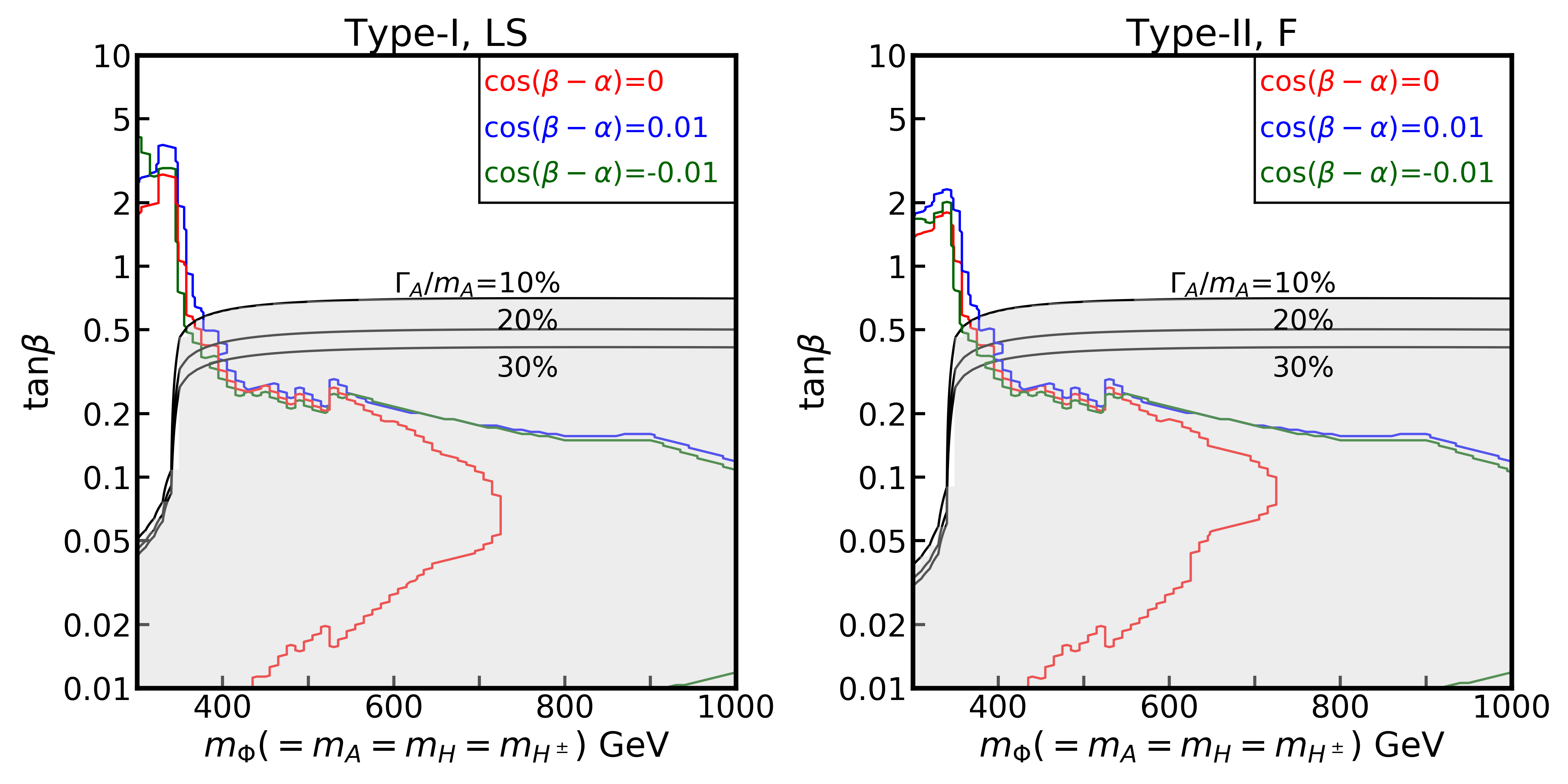}
\caption{Study in the $m_\phi$-$\tan \beta$ plane. Here we choose the benchmark parameter $\cosba=$ 0 (red line), 0.01 (blue line) -0.01 (green line). The left is for the Type-I, LS 2HDM and right for the Type-II, F 2HDM. The grey background represents the points with $\Gamma_A$ or $\Gamma_H$ larger than the values described in \autoref{sec:LHC_reports}, with three black lines for $\Gamma_A$ = 10\%, 20\%, and 30\%.}
 \label{fig:tbma1}
\end{center}
\end{figure}

In \autoref{fig:total_tanbcba}, the combined results are shown in the $\cosba$-$\tan \beta$ plane. In the Type-I 2HDM scenario (left panel), the allowed region is generally around $\cosba=0$. At tree-level, considering the parameter space where the reported limits are applicable, the allowed regions are $\tanb>8, |\cosba|<0.6$, $\tanb<8, |\cosba|<0.02$ and smaller $\tanb$ with smaller $|\cosba|$. At one-loop level, for the non-blue region, the allowed region at $\tanb>1.8$ is similar, while the small $\tanb$ region is totally excluded, even if $\cosba=0$. The Type-II 2HDM results are displayed in the right panel. The main differences occur at $\cosba>0.3$. At the one-loop level, the region $\tanb<1.2$ around $\cosba=0$ is totally excluded except for the blue region. We keep in mind that the blue regions are not currently detectable because of the large heavy Higgs boson decay width in that region.

Further, the combined results are also shown in the $m_\Phi - \tanb$ plane in \autoref{fig:tbma1}. Here we take $m_{H}^2=m_{12}^2/(s_\beta c_\beta)$ and $\cosba=0, \pm0.01$ as the benchmark parameters. We find that the Type-I and Type-LS models have quite similar results because of their similar $hbb$ coupling, shown in the left panel. For the red line with $\cosba=0$, with current published limits the region $m_\Phi < 2 m_t$ GeV for $\tanb<0.5$ can be constrained, and the sensitivity can be extended up to $\tanb\sim3$ with lower masses except for the grey region where the decay width is too large and limits are not applicable any more. When $\cosba$ deviates from exactly 0, such as $\pm 0.01$ (shown by the blue and green lines), we can see that the constraints in the small $\tanb$ region become stronger than those for $\cosba=0$. 
%
In the right panel, we show the Type-II and Type-FL cases which have similar results. The exclusion limits are also similar to the Type-I and Type LS models, except for the moderately reduced constraints on $\tanb$.

With these combined exclusion regions shown in the $\cosba-\tanb$ and $m_\Phi - \tanb$ planes, we find that loop effects in the considered channels are important in the small $\tanb$ region, especially for the $\cosba=0$ region which is excluded by a loop-level analysis except for the space where the limits are not applicable, while the tree-level analysis has no sensitivity. 
We also find that, even though the loop corrections are usually type-dependent, the difference between loop- and tree-level results becomes relatively type-independent.

\section{Conclusions}
\label{sec:con}
Studies of extensions of the Higgs sector of the SM are a promising way to try and address various theoretical and experimental questions following the discovery of the SM-like Higgs boson. In the framework of 2HDMs, we have interpreted current LHC experimental limits on the cross section times branching ratio of the $A\to Zh, H\to VV$ and $H \to hh$ channels at the one-loop level.  
In previous studies, the limits were reported at tree-level, with no limit for the region around $\cosba=0$ because the couplings are proportional to the parameter $\cosba$. At one-loop level, however, we have shown that these results are modified considerably.

Our results for individual channels were displayed in \autoref{fig:AhZ_tanbcba}-\autoref{fig:Hhh_tanbcba} in the $\cosba - \tanb$ plane, which showed that loop effects can contribute significantly in some regions of parameter space, especially in the small $\tanb$ region with $\cosba\sim0$. Through the combined analysis shown in \autoref{fig:total_tanbcba}, we find that the region around $\cosba=0$ with degenerate heavy Higgs masses $m_\Phi$ is detectable using these channels. Except for the regions of parameter space where current limits are not applicable due to large decay width, $\tanb<1.8(1.2)$ can be excluded for the Type-I(II) models, for  a benchmark point with $m_\Phi=300$ GeV.
The combined results in the $m_\Phi - \tanb$ plane were also shown in \autoref{fig:tbma1}. Generally the sensitive region is $\tanb<4$. For $\cosba=0, \pm0.01$, the sensitive region has $m_\Phi$ values up to 350 GeV. For large $m_\Phi$, the $t\bar t$ decay channel opens, resulting in large heavy Higgs decay widths, and the current reported limits are no longer applicable. 
Our study also shows that the improvement of the sensitivity through loop corrections is approximately type-independent.

\acknowledgments
MW, AGW and WS are supported by the Australian Research Council Discovery Project DP180102209. AGW is further supported by the ARC Centre of Excellence for Particle Physics at the Terascale (CoEPP) (CE110001104) and the Centre for the Subatomic Structure of Matter (CSSM). YW is supported by the Natural Sciences and Engineering Research Council of Canada (NSERC).


\appendix

\section{Coupling Formula}
\label{app:Couplings}
Here are the more detailed equations of \autoref{eq:loop-haz}-\autoref{eq:loop_Hhh} for $m_H=m_A=m_{H^\pm}=\sqrt{m_{12}^2/(s_\beta c_\beta)}\equiv m_\phi$ and $\cosba{}=0$,
\begin{align}
C_{hAZ} &\approx \frac{3e^3(p_h^\mu-p_A^\mu)}{64\pi^2s_W^3c_Wm_W^2}\sum_{f=t,b}\xi_fm_f^2\left(\frac{1}{m_\phi^2-m_h^2}\left(2m_f^2\left(B_0^{hff}-B_0^{\phi ff}\right) + m_h^2 B_1^{hff}-m_\phi^2B_1^{\phi ff}\right) \right.\nonumber\\
&\left.\phantom{\frac{m_f^2}{m_\phi^2-m_h^2}}-B_0^{Zff}-B_1^{\phi ff} - 2m_f^2C_0^{hZ\phi f} - m_h^2C_1^{hZ\phi f} - m_\phi^2C_2^{hZ\phi f}\right)\\
C_{HWW} &\approx \frac{3e^3g^{\mu\nu}}{32\pi^2s_W^3m_W}\sum_{f=t,b}\xi_fm_f^2\left(\frac{1}{m_\phi^2-m_h^2}\left(2m_f^2\left(B_0^{hff}-B_0^{\phi ff}\right) + m_h^2 B_1^{hff}-m_\phi^2B_1^{\phi ff}\right)\right.\nonumber\\
&\left.\phantom{\frac{a_1^2}{c_3^3}}-B_0^{Wf\tilde{f}}-B_1^{\phi ff}-B_0^{\phi ff} + 4 C_{00}^{\phi WW ff\tilde{f}} - (m_f^2 + m_{\tilde{f}}^2 - m_W^2)C_0^{\phi WWff\tilde{f}}\right) \nonumber \\
&+ \frac{3e^3 p_1^\mu p_2^\nu}{32\pi^2s_W^3m_W}\sum_{f=t,b}\xi_fm_f^2\left(C_0^{\phi WWff\tilde{f}}+C_2^{\phi WWff\tilde{f}} \right. \nonumber \\
&+4\left.\left(C_1^{\phi WW ff\tilde{f}} + C_{11}^{\phi WWff\tilde{f}} + C_{12}^{\phi WWff\tilde{f}}\right)\right) \\
C_{HZZ} &\approx \frac{3e^3g^{\mu\nu}}{8\pi^2s_W^3c_W^2m_W}\sum_{f=t,b}\xi_fm_f^2\left(\left(\left(c_L^f\right)^2+\left(c_R^f\right)^2\right)\left(-2B_0^{Zff}-B_1^{\phi bb}\phantom{\frac{1}{9}}\right.\right.\nonumber\\
&+\frac{1}{m_\phi^2-m_h^2}\left(2m_f^2\left(B_0^{hff}-B_0^{\phi ff}\right)+m_h^2B_1^{hff}-m_\phi^2B_1^{\phi ff}\right)\nonumber\\
&\left.\left.-(3m_\phi^2+2m_f^2)C_0^{ZZ\phi f}+\frac{1}{9}C_{00}^{ZZ\phi f}-\frac{4m_Z^2+m_\phi^2}{2}C_1^{ZZ\phi f} - 2m_\phi^2C_2^{ZZ\phi f}\right)\right.\nonumber \\
&+c_L^fc_R^f\left(\frac{2}{m_\phi^2-m_h^2}\left(2m_f^2\left(B_0^{\phi ff}-B_0^{hff}\right)+m_\phi^2B_1^{\phi ff}-m_h^2B_1^{hff}\right)\right.\nonumber\\
&\left.\left.\phantom{\frac{1^1}{2^2}}+2B_0^{Zff}+2B_1^{\phi ff} + 4m_f^2C_0^{ZZ\phi f} + m_\phi^2C_1^{ZZ\phi f} +2m_\phi^2C_2^{ZZ\phi f}\right)\right)\nonumber\\
&+\frac{3e^3p_1^\mu p_2^\nu}{8\pi^2s_W^3c_W^2m_W}\sum_{f=t,b}\xi_fm_f^2\left(-2c_L^fc_R^fC_1^{ZZ\phi f}\right.\nonumber\\
&\left.+\left(\left(c_L^f\right)^2+\left(c_R^f\right)^2\right)\left(C_0^{ZZ\phi f}+C_1^{ZZ\phi f}+4\left(C_2^{ZZ\phi f}+C_{12}^{ZZ\phi f}+C_{22}^{ZZ\phi f}\right)\right)\right)\\
C_{Hhh} &\approx \frac{3e^3}{32\pi^2s_W^3m_W^3}\sum_{f=t,b}\xi_fm_f^2\left(\frac{m_b^2(8m_h^2-3m_\phi^2)}{m_\phi^2-m_h^2}B_0^{\phi ff}-\frac{m_b^2(2m_h^2+3m_\phi^2)}{m_\phi^2-m_h^2}B_0^{hff}\right.\nonumber\\
&+\frac{m_\phi^2(8m_h^2-3m_\phi^2)}{2(m_\phi^2-m_h^2)}B_1^{\phi ff}-\frac{m_h^2(2m_h^2+3m_\phi^2)}{2(m_\phi^2-m_h^2)}B_1^{hff} + 6m_f^2B_0^{Zff}+\frac{3m_\phi^2-2m_h^2}{2}B_1^{\phi ff}\nonumber\\
&+m_f^2(8m_f^2+m_h^2+m_\phi^2-m_Z^2)C_0^{hZ\phi f}+2m_f^2(3m_h^2+m_\phi^2-m_Z^2)C_1^{hZ\phi f}\nonumber\\
&\left.\phantom{\frac{1^1}{1^1}}+ 2m_f^2(m_h^2+3m_\phi^2-m_Z^2)C_2^{hZ\phi f}\right)
\end{align}
where $\xi_t = \cot\beta$ for both the Type-I and Type-II models, while $\xi_b = \cot\beta$ for the Type-I model and $\xi_b = -\tan\beta$ for the Type-II model. $\tilde{f}$ is the SU(2) partner of $f$. $B_i^{f_1f_2f_3}\equiv B_i(m_{f_1}^2,m_{f_2}^2,m_{f_3}^2)$, $C_i^{f_1f_2f_3f_4}\equiv C_i(m_{f_1}^2,m_{f_2}^2,m_{f_3}^2,m_{f_4}^2,m_{f_4}^2,m_{f_4}^2)$ and \\ $C_i^{f_1f_2f_3f_4f_5f_6}\equiv C_i(m_{f_1}^2,m_{f_2}^2,m_{f_3}^2,m_{f_4}^2,m_{f_5}^2,m_{f_6}^2)$ are the Passarino-Veltman scalar function~\cite{Passarino:1978jh} in the convention of LoopTools~\cite{Hahn:1998yk}. $c_L^f=I_f - Q_f s_W^2$ and $c_R^f=-Q_f s_W^2$.

\newpage
\clearpage

\bibliographystyle{JHEP}
\bibliography{ref_2hdm13tev}

\end{document}